\documentclass[runningheads]{llncs}
\usepackage[T1]{fontenc}
\usepackage{graphicx}
\begin{document}
\title{Coarse-to-Fine Joint Registration of MR and Ultrasound Images via Imaging Style Transfer
}
\author{Junyi Wang\inst{1} \and
Xi Zhu\inst{1} \and Yikun Guo\inst{1} \and Zixi Wang\inst{1} \and Haichuan Gao\inst{1} \and Le Zhang\inst{1} \and Fan Zhang\inst{1,*}}
\authorrunning{J. Wang et al.}
\institute{University of Electronic Science and Technology of China, Chengdu, Sichuan, China
\\\email{fan.zhang@uestc.edu.cn}
}
\maketitle              % typeset the header of the contribution
\begin{abstract}
We develope a pipeline for registering pre-surgery Magnetic Resonance (MR) images and post-resection Ultrasound (US) images. Our approach leverages unpaired style transfer using 3D CycleGAN to generate synthetic T1 images, thereby enhancing registration performance. Additionally, our registration process employs both affine and local deformable transformations for a coarse-to-fine registration. The results demonstrate that our approach improves the consistency between MR and US image pairs in most cases.

\end{abstract}
\section{Introduction}
Magnetic resonance imaging (MRI) is widely used in planning brain tumor resection, providing a comprehensive view of brain tissues. However, acquiring MR images is expensive, time-consuming, and environmentally constrained. Ultrasound (US) imaging is another popular method for image guidance during surgery \cite{Taylor1996-ga}, though it is limited in resolution and contrast. Thus, the registration of pre-surgery MR images to intra-surgery ultrasound images can provide more information for tissues, which can be crucial for effective surgical planning \cite{Juvekar2024-kh}.

One of the main goals is to align images from different modalities into the same space \cite{Balakrishnan2019-mn}. This requires estimating the shift and rotation to register between the MR image space and the US image space. Additionally, since the US image is acquired after tumor resection, there can be local deformations around the resected tissues \cite{Heinrich2013-lz,Maurer1998-ju}, necessitating advanced estimation for precise alignment.

Unlike monomodality registration, the signal distribution in US and MR images is quite different. Therefore, there is a need to use a modality-invariant metric to quantify the similarity between the two images and accurately identify the corresponding tissues \cite{Heinrich2013-lz,Heinrich2012-au,Liu2022-nd}. This presents the main challenge for multimodality registration. In this work, we employed image style transfer techniques: CycleGAN to achieve a unified distribution of signals \cite{Yang2018-wa,Zhu2017-ss}. Additionally, in the registration process, we utilized block matching approach \cite{Modat2014-nq,Ourselin2001-ri} for affine transform estimation, and a pretrained SynthMorph model \cite{Hoffmann2022-ds} for the inference of local deformation. 

For the subtask ReMIND2Reg in the 2024 Learn2Reg challenge, we developed a pipeline for estimating both the global brain rotation and shifts, as well as the local deformation. 

\section{Methods}
\begin{figure}
    \centering
    \includegraphics[width=\textwidth]{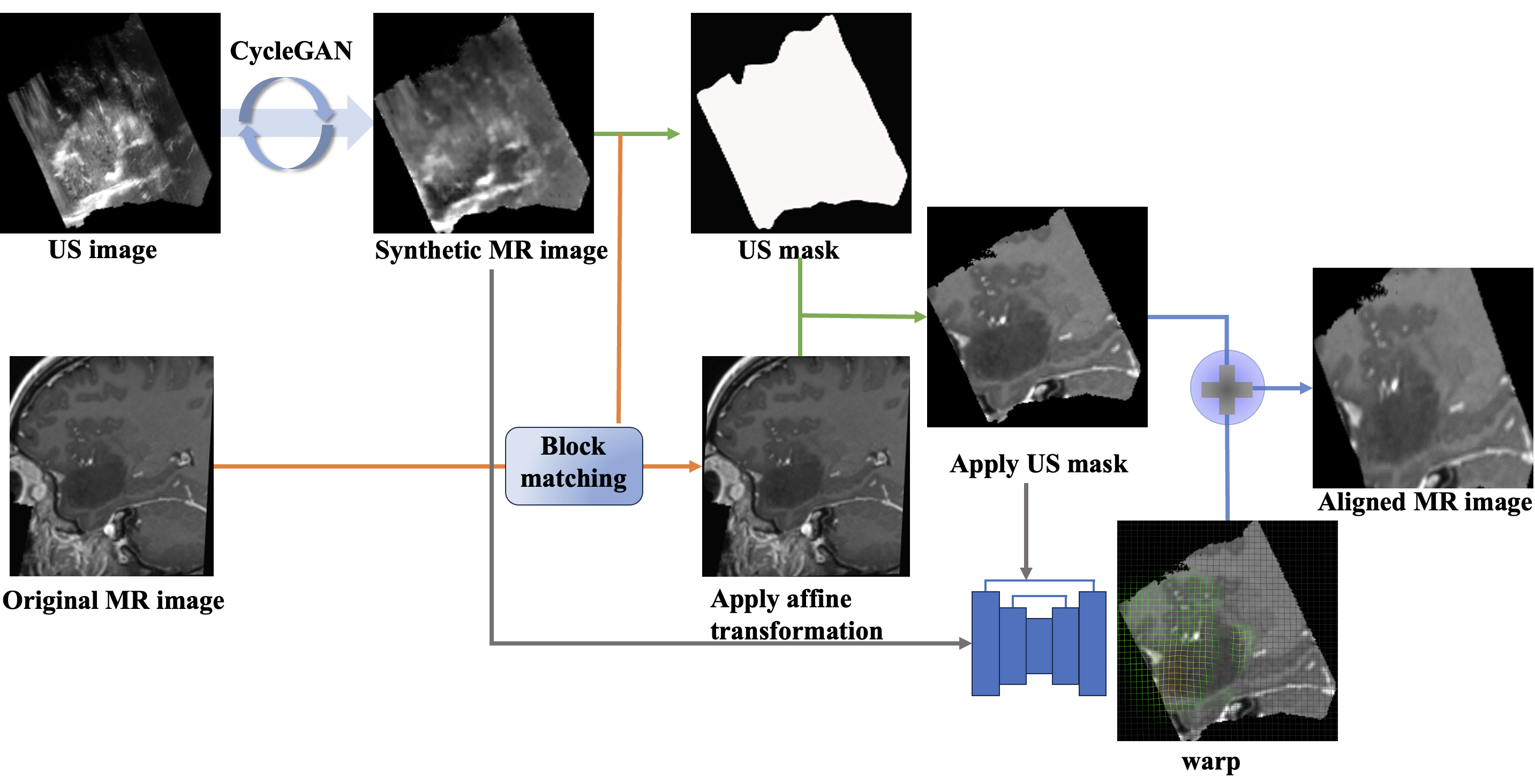}
    \caption{Pipeline for the registration of MR image acquired pre surgery and US image after tumor resection.}
    \label{fig:1}
\end{figure}

In this work, we develop a novel approach for registration between Ultrasound image and MR image, as illustrated in Fig. \ref{fig:1}. Our approach can be divided into 3 phases: 1. Brain imaging style transfer. 2. affine transformation with block matching 3. local deformation estimation.

\subsection{Brain imaging style transfer}
In this stage, we employed CycleGAN to generate T1-style images from Ultrasound images. This method is inspired by the most advanced registration techniques developed for T1 pairs \cite{Balakrishnan2019-mn,Chen2022-uh} and the proven success of CycleGAN in style transfer \cite{Li2023-uo,Yang2018-wa,Zhu2017-ss}. Unlike other generation methods, CycleGAN does not require paired images for training, making it particularly useful for generageneratingting data when paired images are unavailable. In the challenge dataset, although each subject has both modalities, the data cannot be considered paired because there is no strict registration between the modalities. Moreover, during the inference process, due to the computation characteristics of CNN, CycleGAN relies solely on local information to produce ideal images, thereby preventing the generation of artifacts or tissue shifts in the resulting images. This localized approach enhances the reliability of subsequent image registration.

\begin{figure}
    \centering
    \includegraphics[width=\textwidth]{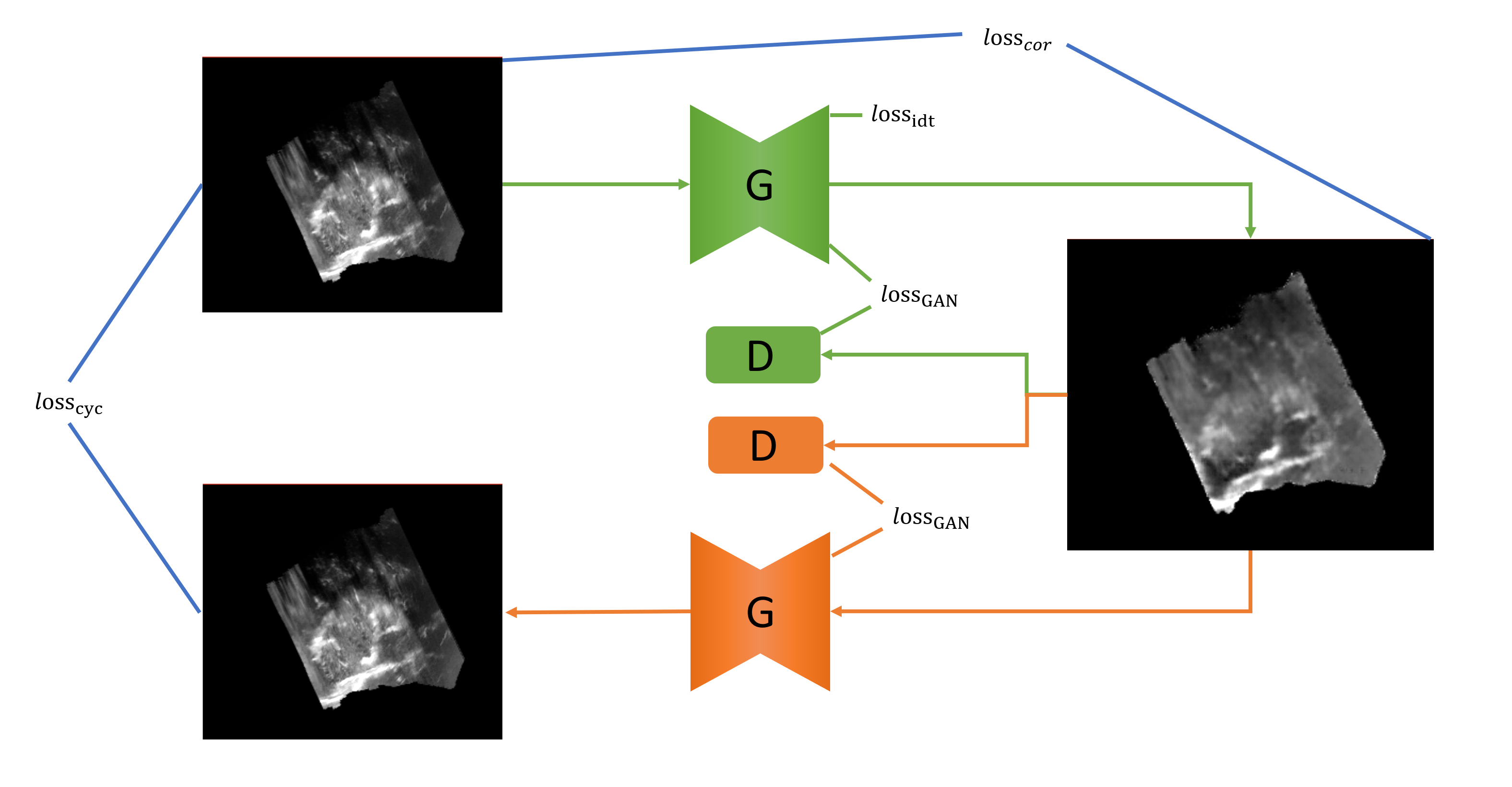}
    \caption{Training process of CycleGAN.}
    \label{fig:2}
\end{figure}

Therefore, we choose 3D-CycleGAN \cite{Sun2023-bm} to complete the conversion process between the US images and T1w MRI, which can retain more spatial information compared with 2D ones. As shown in Fig. \ref{fig:2}, the entire model is divided into two generating processes and each of them has generator and discriminator. For generators, we used U-net which downsampling depth is 7 as the backbone. Loss function is composed of several parts:

$$L= loss_{GAN}+loss_{cyc}+loss_{idt}+loss_{cor}$$

$loss_{GAN}$ aims to optimize generators to create realistic fake data, while discriminators become better at distinguishing fake data from real data. $loss_{cyc}$ aims to measure the difference between the generated fake source image and the input source image when the generated target fake image is used as input. $loss_{idt}$ is designed so that when the input of the generator is the target domain image, the output image is consistent with the input. $loss_{cor}$ measures the correlation between the input source image and generated fake image to prevent generators from generating false structures.
For the T2 image, we utilized the advanced developed technique T1ify to generate a  similar contrast profile as T1. The subsequent process is the same as T1 images.

\subsection{ Estimation of affine transformation}
For the affine registration process we utilize hierarchical block matching for an optimal affine matrix transformation as suggested in \cite{Zhou2019-ea}. On each level, we identify the corresponding points in both moving and fixed images iteratively, and then estimate the affine parameters through least trimmed squares regression (LTS). Notably the proportion of points considered in the regression is one of the hyperparameters, which needs a grid search process for the best performance.
The selection of corresponding points is based on the partition of the image. In our approach each of the images is divided into uniform blocks with 4 voxel size and the 25\% blocks with highest intensity variance are subsequently used in the matching process. We compare the absolute normalized cross correlation between each block in the moving image and all blocks in the fixed image to find the best match pairs as the corresponding points.

The affine registration algorithm using block matching in this pipeline utilizes the official implementation of the open-source software NiftyReg \cite{Modat2014-nq}. 

\subsection{Estimation of local deformation}
For local deformation estimation, we used a pretrained SynthMorph model to generate the deformable field. Given that the local deformation stem from tumor resection \cite{Juvekar2024-kh} while the other regions remain stationary, we crop the moving T1 images after affine alignment with post-resection ultrasound images. This is done with the assumption that no deformation exists in the unavailable areas with ultrasound, as indicated by the mask. 

SynthMorph \cite{Hoffmann2022-ds}, a variant of VoxelMorph \cite{Balakrishnan2019-mn}, belongs to a series of representative deep learning methods for registration tasks in different modalities \cite{Balakrishnan2019-mn,Zhang2022-ko,Miyake2022-gx}and different tissues \cite{Heinrich2022-bl,Miyake2022-gx}. It is trained on diverse synthetic label maps and images using a loss function that penalizes shape differences, ensuring that the model is not constrained to a specific modality and can generalize effectively. This approach has demonstrated superior performance in multimodality and tumor existing instances \cite{Hoffmann2024-ii}. 

\subsection{Experiments}
There are 97 subjects in the dataset that both have ultrasound images and T1w MRI. We selected 80 subjects’ data for CycleGAN training, remaining 17 subjects for testing, For the hyperparameters, we used AdamW optimizer with a learning rate of  and set training epochs to 400.  All computation was conducted on RTX 3090 GPUs.

\begin{figure}
    \centering
    \includegraphics[width=0.7\textwidth]{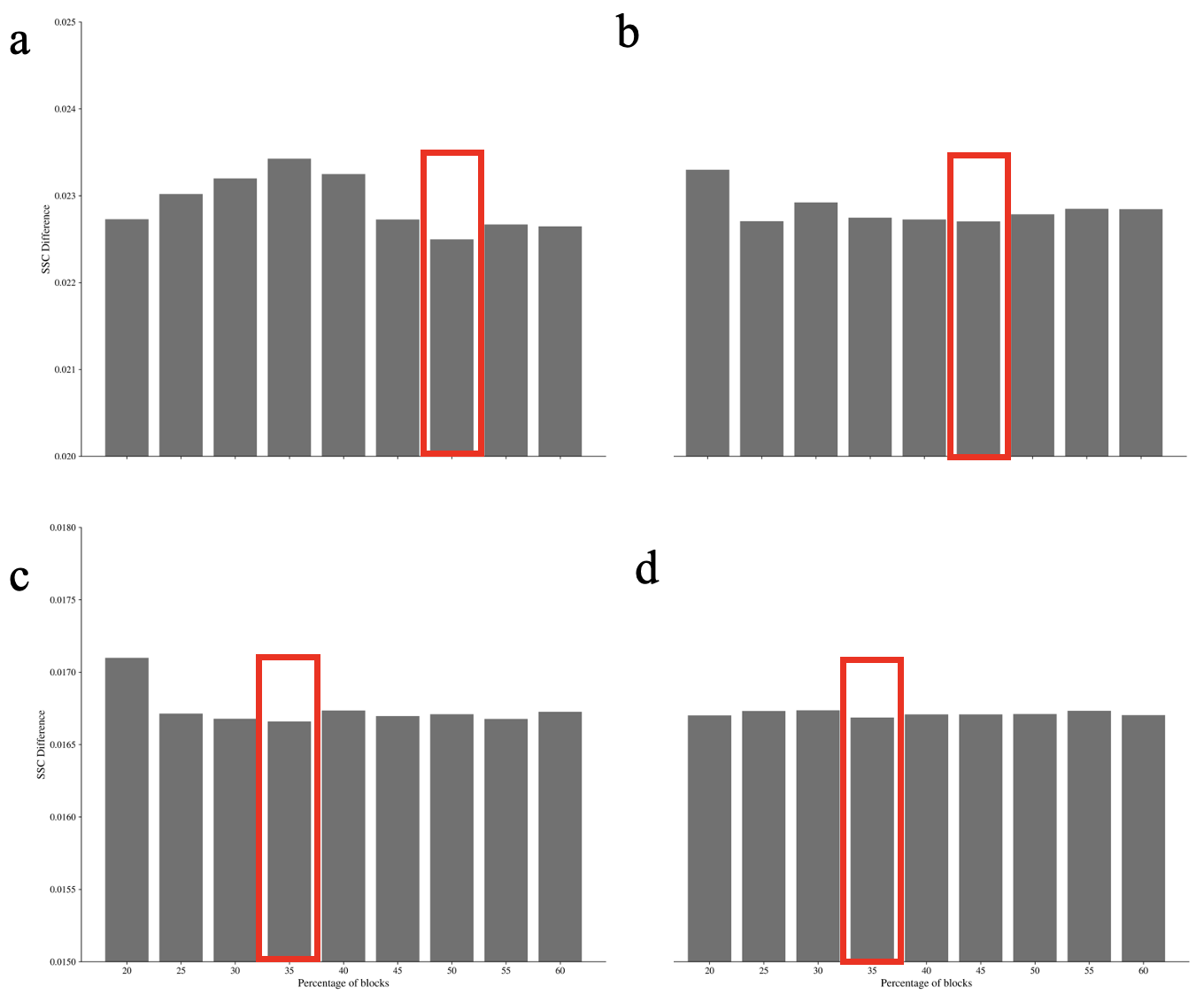}
    \caption{The MSEs of SSC on 4 subjects on validation set for different proportions of blocks considered during the LTS process. The bar framed by a red rectangle represents the minimal difference achieved, which is the metric we used for selecting the number of blocks.}
    \label{fig:3}
\end{figure}

In the affine registration process, utilizing the NiftyReg software for block matching, we employ a two-level registration approach. The first level is performed at a 1 mm resolution, followed by a second level at a 0.5 mm resolution to obtain a more precise affine transformation estimation. At each level, we divide the image into blocks of 4 voxels, and the 25\% of blocks with the highest intensity variance are selected for the matching process. The matching process begins at the centers of both images, assuming they are approximately aligned in a common space. After a grid search to optimize performance, we use the optimal proportion of corresponding blocks in the LTS process for each subject according to the context difference, as shown in Fig. \ref{fig:3}.

The grid search process utilized self-similarity context (SSC) difference as the metric \cite{Heinrich2013-lz}, This metric has demonstrated robustness in describing image context. Therefore, in this work, we compute the mean squared error(MSE) of the SSC of each image pair in the validation set to select the optimal parameters.

In the final local deformation estimation, we utilize the official implementation of SynthMorph \cite{Hoffmann2024-ii} as a sub-module in the FreeSurfer software. We set the smoothness parameter to 0.5 to balance registration precision and warp field smoothness.

\section{Results}
\begin{figure}
    \centering
    \includegraphics[width=0.7\textwidth]{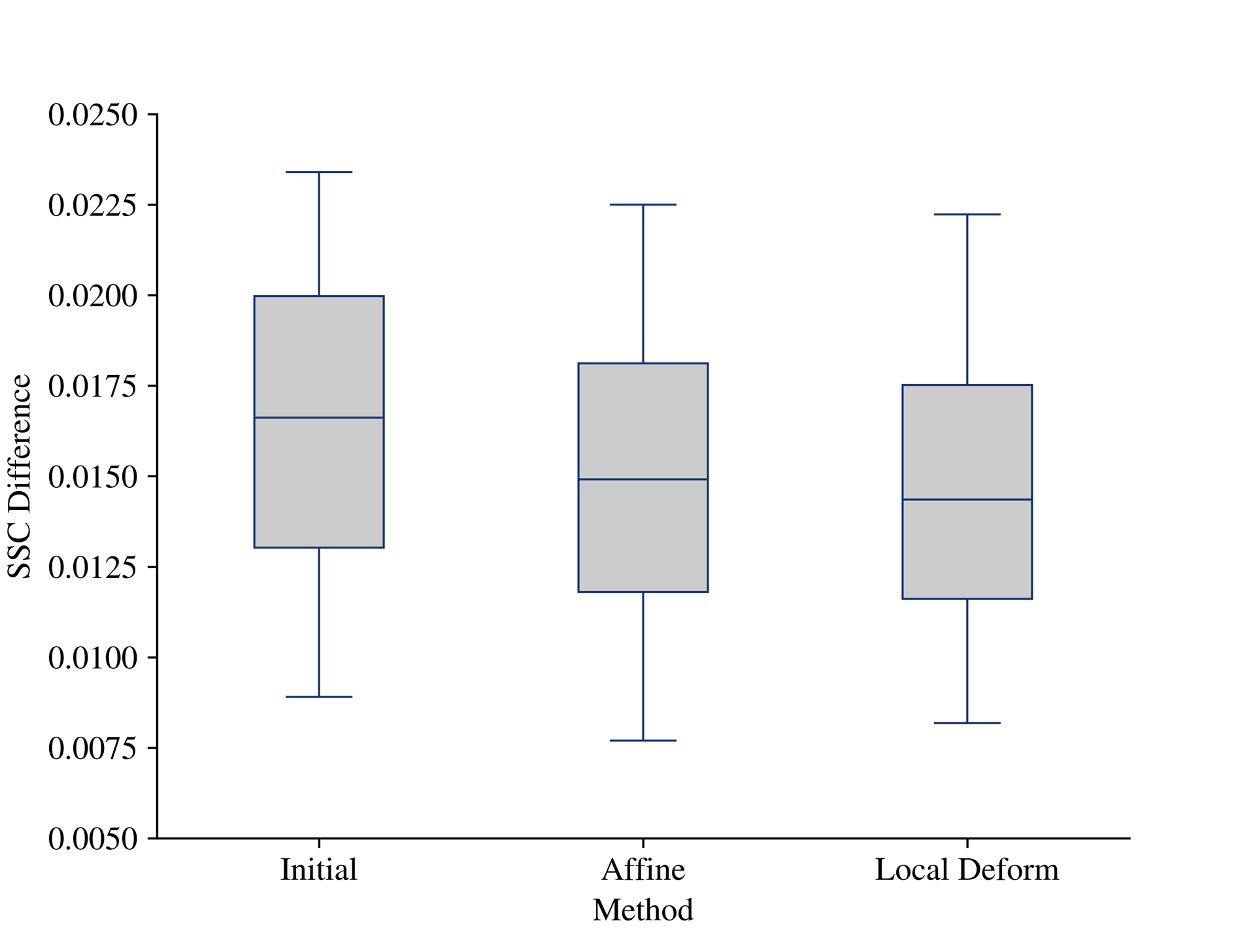}
    \caption{MSEs of SSC on validation set after different phases.}
    \label{fig:4}
\end{figure}

\begin{figure}
    \centering
    \includegraphics[width=\textwidth]{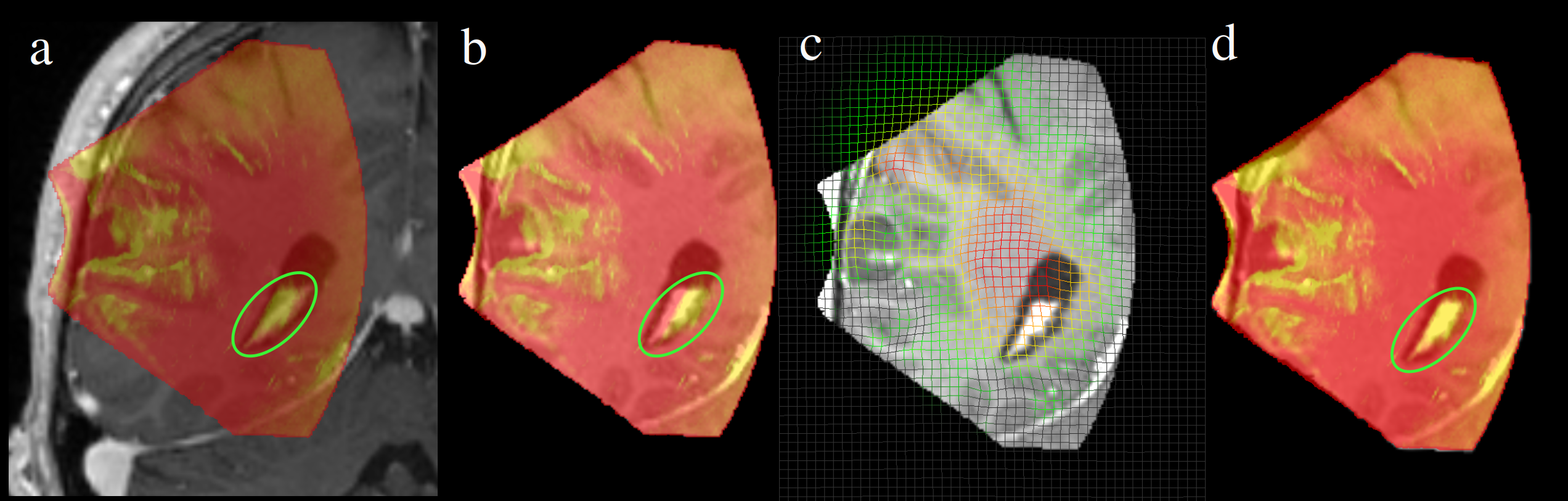}
    \caption{The visualization of the deformation field, and the overlap of fixed US image and MR images after different stages. a) The difference of original MR image with US image, b) The difference of MR image after an affine alignment and US image c) deformable field applied on MR image d) The difference of final MR image and US image.}
    \label{fig:5}
\end{figure}

Compared to the original image pair difference, the application of both the affine transform and local deformation results in better alignment, as evidenced by a decrease in the mean squared error (MSE) of the self-similarity context (SSC)(Fig. \ref{fig:4}).
However, this metric shows inadequate consistency with the Target Registration Error (TRE). In some cases, a decrease in the MSE of SSC does not correspond to a similar reduction in TRE. Although the SynthMorph shows more consistency with fixed US image in the circled area, as depicted in Fig. \ref{fig:5}, the TRE rises from 1.60 mm to 2.38 mm. 
\section{Discussion}
In this work, we developed an automated registration approach to align pre-surgery MR images with post-tumor resection US images. Our method effectively estimates brain shifts through a block-matching approach and infers local deformations around the resection areas. Although this approach achieves a more consistent self-similarity context, it sometimes results in implausible alignments with higher Target Registration Errors (TRE). Therefore, identifying a context description metric with greater consistency with TRE is crucial. Additionally, examining the discrepancies between SynthMorph registration results and misaligned landmarks will provide further insights for improvement.

\bibliographystyle{splncs04}
\bibliography{References}

\begin{thebibliography}{10}
\providecommand{\url}[1]{\texttt{#1}}
\providecommand{\urlprefix}{URL }
\providecommand{\doi}[1]{https://doi.org/#1}

\bibitem{Balakrishnan2019-mn}
Balakrishnan, G., Zhao, A., Sabuncu, M.R., Guttag, J., Dalca, A.V.: {VoxelMorph}: A learning framework for deformable medical image registration. IEEE Trans. Med. Imaging  \textbf{38}(8),  1788--1800 (Feb 2019)

\bibitem{Chen2022-uh}
Chen, J., Frey, E.C., He, Y., Segars, W.P., Li, Y., Du, Y.: {TransMorph}: Transformer for unsupervised medical image registration. Med. Image Anal.  \textbf{82},  102615 (Nov 2022)

\bibitem{Heinrich2022-bl}
Heinrich, M.P., Hansen, L.: Voxelmorph++. Biomedical Image Registration pp. 85--95 (2022)

\bibitem{Heinrich2012-au}
Heinrich, M.P., Jenkinson, M., Bhushan, M., Matin, T., Gleeson, F.V., Brady, S.M., Schnabel, J.A.: {MIND}: modality independent neighbourhood descriptor for multi-modal deformable registration. Med. Image Anal.  \textbf{16}(7),  1423--1435 (Oct 2012)

\bibitem{Heinrich2013-lz}
Heinrich, M.P., Jenkinson, M., Papiez, B.W., Brady, S.M., Schnabel, J.A.: Towards realtime multimodal fusion for image-guided interventions using self-similarities. Med. Image Comput. Comput. Assist. Interv.  \textbf{16}(Pt 1),  187--194 (2013)

\bibitem{Hoffmann2022-ds}
Hoffmann, M., Billot, B., Greve, D.N., Iglesias, J.E., Fischl, B., Dalca, A.V.: {SynthMorph}: Learning {Contrast-Invariant} registration without acquired images. IEEE Trans. Med. Imaging  \textbf{41}(3),  543--558 (Mar 2022)

\bibitem{Hoffmann2024-ii}
Hoffmann, M., Hoopes, A., Greve, D.N., Fischl, B., Dalca, A.V.: Anatomy-aware and acquisition-agnostic joint registration with {SynthMorph}. Imaging Neurosci (Camb)  \textbf{2},  1--33 (Jun 2024)

\bibitem{Juvekar2024-kh}
Juvekar, P., Dorent, R., K{\"o}gl, F., Torio, E., Barr, C., Rigolo, L., Galvin, C., Jowkar, N., Kazi, A., Haouchine, N., Cheema, H., Navab, N., Pieper, S., Wells, W.M., Bi, W.L., Golby, A., Frisken, S., Kapur, T.: {ReMIND}: The brain resection multimodal imaging database. Sci Data  \textbf{11}(1), ~494 (May 2024)

\bibitem{Li2023-uo}
Li, Z., Shang, Z., Liu, J., Zhen, H., Zhu, E., Zhong, S., Sturgess, R.N., Zhou, Y., Hu, X., Zhao, X., Wu, Y., Li, P., Lin, R., Ren, J.: {D-LMBmap}: a fully automated deep-learning pipeline for whole-brain profiling of neural circuitry. Nat. Methods  \textbf{20}(10),  1593--1604 (Oct 2023)

\bibitem{Liu2022-nd}
Liu, S., Yang, B., Wang, Y., Tian, J., Yin, L., Zheng, W.: {2D/3D} multimode medical image registration based on normalized {Cross-Correlation}. NATO Adv. Sci. Inst. Ser. E Appl. Sci.  \textbf{12}(6), ~2828 (Mar 2022)

\bibitem{Maurer1998-ju}
Maurer, Jr, C.R., Hill, D.L., Martin, A.J., Liu, H., McCue, M., Rueckert, D., Lloret, D., Hall, W.A., Maxwell, R.E., Hawkes, D.J., Truwit, C.L.: Investigation of intraoperative brain deformation using a 1.5-t interventional {MR} system: preliminary results. IEEE Trans. Med. Imaging  \textbf{17}(5),  817--825 (Oct 1998)

\bibitem{Miyake2022-gx}
Miyake, N., Lu, H., Kamiya, T., Aoki, T., Kido, S.: Temporal subtraction technique for thoracic {MDCT} based on residual {VoxelMorph}. NATO Adv. Sci. Inst. Ser. E Appl. Sci.  \textbf{12}(17), ~8542 (Aug 2022)

\bibitem{Modat2014-nq}
Modat, M., Cash, D.M., Daga, P., Winston, G.P., Duncan, J.S., Ourselin, S.: Global image registration using a symmetric block-matching approach. J Med Imaging (Bellingham)  \textbf{1}(2),  024003 (Jul 2014)

\bibitem{Ourselin2001-ri}
Ourselin, S., Roche, A., Subsol, G., Pennec, X., Ayache, N.: Reconstructing a {3D} structure from serial histological sections. Image Vis. Comput.  \textbf{19}(1),  25--31 (Jan 2001)

\bibitem{Sun2023-bm}
Sun, B., Jia, S., Jiang, X., Jia, F.: Double {U-Net} {CycleGAN} for {3D} {MR} to {CT} image synthesis. Int. J. Comput. Assist. Radiol. Surg.  \textbf{18}(1),  149--156 (Jan 2023)

\bibitem{Taylor1996-ga}
Taylor, R.H.: Computer-integrated Surgery: Technology and Clinical Applications. MIT Press (1996)

\bibitem{Yang2018-wa}
Yang, H., Sun, J., Carass, A., Zhao, C., Lee, J., Xu, Z., Prince, J.: Unpaired brain {MR-to-CT} synthesis using a {Structure-Constrained} {CycleGAN}. In: Deep Learning in Medical Image Analysis and Multimodal Learning for Clinical Decision Support. pp. 174--182. Springer International Publishing (2018)

\bibitem{Zhang2022-ko}
Zhang, F., Wells, W.M., O'Donnell, L.J.: Deep diffusion {MRI} registration ({DDMReg)}: A deep learning method for diffusion {MRI} registration. IEEE Trans. Med. Imaging  \textbf{41}(6),  1454--1467 (Jun 2022)

\bibitem{Zhou2019-ea}
Zhou, L., Heller, N., Shi, Y., Xiao, Y., Sznitman, R., Cheplygina, V., Mateus, D., Trucco, E., Sharon~Hu, X., Chen, D., Chabanas, M., Rivaz, H., Reinertsen, I.: {Large-Scale} Annotation of Biomedical Data and Expert Label Synthesis and Hardware Aware Learning for Medical Imaging and Computer Assisted Intervention: International Workshops, {LABELS} 2019, {HAL-MICCAI} 2019, and {CuRIOUS} 2019, Held in Conjunction with {MICCAI} 2019, Shenzhen, China, October 13 and 17, 2019, Proceedings. Springer Nature (Nov 2019)

\bibitem{Zhu2017-ss}
Zhu, J.Y., Park, T., Isola, P., Efros, A.A.: Unpaired image-to-image translation using cycle-consistent adversarial networks. ICCV pp. 2242--2251 (Mar 2017)

\end{thebibliography}




% LNCS Homepage, \url{http://www.springer.com/lncs}, last accessed 2023/10/25
% \end{thebibliography}
\end{document}